\documentclass[preprint,12pt,authoryear]{elsarticle}

\usepackage{amssymb}
\usepackage{hyperref}
\usepackage[nointegrals]{wasysym}
\usepackage{hyperref}
\usepackage{xcolor}
\usepackage{amsmath}

\def\HypCat{{\it Hypatia Catalog }}
\def\Hyp{{\it Hypatia }}

\journal{Treatise on Geochemistry, 3$^{rd}$ Ed.}

\def\aj{{AJ}}
\def\apj{{ApJ}}
\def\apjs{{ApJS}}
\def\apjl{{ApJL}}
\def\araa{{ARA\&A}}
\def\aap{{A\&A}}
\def\pasp{{PASP}}

\def\mnras{{MNRAS}}

\def\icarus{{Icarus}}

\begin{document}

\begin{frontmatter}



\title{Abundances of Elements in Solar Systems}


\author[inst1]{Natalie R. Hinkel}

\affiliation[inst1]{organization={Louisiana State University},
 addressline={Department of Physics \& Astronomy}, 
 city={Baton Rouge}, 
 state={LA},
 postcode={70803},
 country={USA}}

\author[inst2]{Edward D. Young}

\affiliation[inst2]{organization={University of California, Los Angeles},
 addressline={Department of Earth, Planetary, \& Space Sciences}, 
 city={Los Angeles}, 
 state={CA},
 postcode={90095},
 country={USA}}

\begin{abstract}

The relationship between stars and planets provides important information for understanding the interior composition, mineralogy, and overall classification of small planets (R $\lesssim$ 3.5 R$_{\oplus}$). Since stars and planets are formed at the same time and from the same material, their compositions are inextricably linked to one another, especially with respect refractory elements like Mg, Si, and Fe. As a result, stellar elemental abundances can help break the degeneracy inherent to planetary mass-radius models and determine whether planets may be similar to the Earth in composition or if additional factors, such as formation near the host star or a giant impact, may have influenced the planet's make-up. To this end, we now have observations of the abundances of extrasolar rocks that were pulled onto the surfaces of a white dwarfs, whose compositions act as a direct insight into the interiors of small exoplanets. From measurements of $\sim$30 of these ``polluted" white dwarfs, we have found that composition of the extrasolar rocks are similar to Solar System chondritic meteorites.

\end{abstract}



\begin{keyword}
exoplanet \sep interior composition \sep mini-Neptune \sep planet classification \sep polluted white dwarf \sep refractory elements \sep solar neighborhood \sep stellar abundances \sep stellar spectroscopy \sep super-Earth  \sep super-Mercury 
\PACS 0000 \sep 1111
\MSC 0000 \sep 1111
\end{keyword}

\end{frontmatter}



\noindent
\textbf{Key Points:}
\begin{itemize}
    \item The mass-radius relationship for exoplanets provides degenerate results for interior composition \vspace{-2mm}
    \item Stellar elemental abundances for Mg, Si, and Fe can break the degeneracy due to the compositional connection between stars and their planets\vspace{-2mm}
    \item Physical/chemical processes may have influenced planet formation, creating planets that are denser (e.g., super-Mercury) or less dense (e.g., mini-Neptune) than expected from stellar abundances \vspace{-2mm}
    \item Utilizing stellar abundances for planet interiors requires an understanding of the chemistry within the solar neighborhood (via \HypCat and large stellar surveys) and the variations between different abundance methodologies \vspace{-2mm}
    \item White dwarf stars that have pulled rocky material onto their surface (i.e. that are ``polluted") provide a means to observe the interior composition of extrasolar rocks\vspace{-2mm}
    \item The composition of material on $\sim$30 observed polluted white dwarfs is very similar to CI chondrites, bulk Earth, and the overall make-up of the Solar System    \vspace{-2mm}
\end{itemize}

\section{Introduction}
The compositions of extrasolar rocky planets provide context for the terrestrial planets of our Solar System. Understanding the interior composition of an exoplanet is also essential to fully characterizing the planet and to determining whether it can support life. In short, we wish to know whether Earth and the other small, rocky bodies (R $\lesssim$ 3.5 R$_{\oplus}$) \citep{Bergsten22} orbiting the Sun are typical or somehow unusual. With the plethora of new data for exoplanets, progress is being made. Per current observation techniques\footnote{Statistics from the NASA Exoplanet Archive (Jan 2023) \url{https://exoplanetarchive.ipac.caltech.edu/}}, $\sim$76\% of all known planets have radius measurements, $\sim$44\% have mass measurements, and $\sim$20\% have both mass and radius measurements. When it comes to smaller, potentially Earth-like planets with R$_{\oplus} < 2.0$ \citep{Fulton17}, which make up $\sim$31\% of the total, only $\sim$10\% of the small planets (or $\sim$3\% of the total) have both mass and radius measurements. Yet it's through the planetary mass-radius relationship that bulk densities are determined, which can be inverted for relative abundances of iron-rich metal cores and silicate mantles, ultimately defining the mineralogy and interior structure of small planets. Some of the first mass-radius models incorporated a solid Fe-core and mantle \citep[e.g.,][]{Valencia06, Seager07}, later expanding to include a volatile layer on the surface \citep[e.g.,][]{Zeng13, Zeng16}. However, results from these models produced significant degeneracies, such that multiple combinations of different sized layers could reproduce the observed bulk planetary densities. Even more sophisticated thermodynamic interior models, which are able to produce accurate mineralogies for small planets \citep[e.g.,][]{Dorn15,Unterborn16, Unterborn23}, have problems with degeneracies. Fortunately, many of these degeneracies can be broken by virtue of the fact that stars and planets are formed at the same time from the same molecular cloud, such that the amount of certain elements, or abundances, comprising the planet are reflected in the composition of the host star. However, this methodology does not work for all planetary scenarios.

In this chapter, we discuss the chemical make-up of planets using the compositional relationship between stars and their planets. In \S \ref{s.interplay}, we first describe the standard way that the abundances of elements within stars are measured. We then explain those instances where the 1-to-1 chemical relationship between a star and planet has been tested and verified. Finally, we explain in which scenarios additional physical or chemical processes may need to be invoked in order to understand a planet's interior. In \S \ref{s.neighborhood}, we discuss the composition of stars within the local solar neighborhood, focusing specifically on the \HypCat database of stellar abundances as well as large stellar abundance surveys. Given the large amount of elemental abundance data for nearby stars (and their planets), we briefly describe the difference between various spectroscopic techniques and the ways in which they may impact resulting abundance determinations. Finally, in \S \ref{s.wd}, we explain how the accretion of rocky bodies (e.g., asteroids, comets, moons, or planets) onto the surface of white dwarfs provides a distinct but complementary window into the composition of extrasolar bodies. 

%

\section{The Chemical Interplay Within a Planetary System}\label{s.interplay}

In order to determine the ways in which the make-up of planets may reflect that of their host star, we must first understand how stellar elemental abundances are determined and what they mean within a planetary context. We will then go over the compositional interplay between stars and exoplanets, when it succeeds but also when additional chemical or physical mechanisms may be necessary to explain if/when the composition of the planet differs from the host star. 

\subsection{Measuring the Compositions of Stars}
The Big Bang only produced H, He, and a small amount of Li at the beginning of the Universe. With the exception of Li, Be, and B that are made via spallation (or when high energy ``particles" like cosmic rays impact matter), the remaining elements within the Periodic Table were created during the life and death of multiple generations of stars. For example, elements were produced within stellar interiors, during stellar explosions (i.e. supernovae), as a result of stellar mergers, or during the final death throes of small stars (such as asymptotic giant branch stars, or AGBs). As a result, large molecular clouds, from which stars are formed, contains elements produced from a wide variety of stellar sources over time. 

Astronomers determine the composition of a star by measuring its flux over a wavelength range, which results in a spectrum. Depending on the star and/or wavelength regime (e.g., ultraviolet, optical, infrared, etc.), notable lines of lower intensity in the star's spectrum correlate with atomic (and sometimes molecular) absorption features. The strength of the line (width and depth) indicates the number of atoms (or molecules) that must be present within the upper layers, or photosphere, of the star. This is then used to calculate the total amount, or abundance, of an element (or molecule) within the stellar photosphere. 

In chemistry or geology (particularly geochemistry), the number of atoms present within an object are described using moles (or mols), or perhaps in weight percent. However in stellar spectroscopy, element abundances are reported as $\log_{10}$ and scaled such that the abundance of H $\equiv$ 12. The results are then compared as a ratio to another element -- because elements have different formation mechanisms which happen on unique timescales, and normalized with respect to that same elemental ratio within the Sun (indicated with square [\,] brackets). \citet{Hinkel22} presents an in-depth mathematical framework for deriving stellar abundances and their implications, as well as some historical context and community-driven caveats. Because the underlying math describing stellar elemental abundances is fraught with omissions, multiple normalizations, and underlying assumptions within the literature, this framework helps to clarify the details to facilitate understanding and conversion to molar fractions. While most of these explanations are beyond the scope of this discussion, it's important to understand that stellar abundances are typically presented as [Q/H], where Q is a general element and H hydrogren, the most abundant element within the Universe. The implicit solar normalization allows comparisons to the Sun (which is observed often but not always ``well-measured", see \S 6 of \citealt{Hinkel22}). In this scheme, a stellar relative abundance value of [Q/H] = 0.0 dex indicates that the ratio of Q/H in the star is the same as that in the Sun. The dex unit, meaning ``decadic logarithmic unit" \citep{Lodders19}, is a base-10 logarithm, similar to a decibel \citep[or dB, see][for more details]{Hinkel22}. The dex unit is useful to astronomers not only because it is helpful when tracking observations that span a large dynamic range, but also because it allows for an inherent comparison to the Sun's composition to understand whether a star is super- or sub-solar in its respective abundances.

\subsection{The Compositional Link Between Stars and Their Planets }

The photosphere of a star (other than the Sun) is the only region that astronomers are able to observe to determine composition. Despite the fact that it is a small fraction of the star, the photosphere is nonetheless extremely valuable because it hasn't been strongly impacted by processes within the stellar interior \citep[although many processes are time-dependent so this is less true for old stars,][]{Lodders09, Lodders19}. In general, though, the composition of the stellar photosphere is very similar to the composition of the molecular cloud from which it -- as well as orbiting planets, moons, asteroids, etc. -- originally formed. For example, within the Solar System, the Sun, Earth, and Mars all have the same relative proportions of the major rocky planet building elements, e.g., Mg, Si, and Fe, to within 10\% \citep{Wanke94, mcdonough03, lodders2003, Unterborn19, Unterborn23}. It is therefore possible to use our understanding of the make-up of stars in order to infer the composition of other bodies within their planetary systems, especially small planets that are predominantly composed of Mg, Si, and Fe (in addition to O, which combines with the elements to create minerals). To this end, there are a variety of studies that confirm or rely on the direct (1-to-1) compositional link between stars and their planets, e.g. \citet[e.g.,][]{Lodders03, Santos15, Santos17, Dorn15, Dorn17a, Dorn17b, Hinkel18, Putirka19, Plotnykov20, Putirka21a, Putirka21b}.

\citet{Bond10a} were some of the first to look at the direct compositional connection between stars and planets, which they did by simulating the dynamic and chemical properties during small planet (0.05-1.42 M$_{\oplus}$) formation. They assumed that the material within the circumstellar disk reflected equilibrium condensation from the protoplanetary solar nebula, even as it evolved, and found that the ratios of refractory elements had not significantly changed for the bulk abundances of the small planets, once they formed. In a follow-up paper, \citet{Bond10b} noted that, within small planets, ``compositional variations are produced by variations in the elemental abundances of the host star and thus the system as a whole." A markedly different study by \citet{Thiabaud15} sought to invert the problem and specifically test the chemical relationship between stars and their planets. They modeled the composition of the protoplanetary disk to determine the Fe/Si, Mg/Si, and C/O ratios within rocky, ice, and giant gaseous planets and then compared their composition to the stellar abundances of the host star. For all three types of planets, and for modeled planets with and without irradiation of the planetary disk, the element ratios for Mg, Si, and Fe between stars and planets correlate along a 1-to-1 relationship. Namely, the Fe/Si and Mg/Si ratios within planets -- formed from the stellar nebula -- matched the same ratios within the host star.

Similarly, \citet{Putirka19} used stellar abundances of non-volatile elements from 4,382 stars within the Hypatia Catalog \citep[][see Section \ref{s.hypatia}]{Hinkel14} to test the assumption that planetary and stellar compositions are similar. They determined the bulk silicate composition of theoretical exoplanets in order to estimate interior mineral proportions by mass balancing the major oxides. Using this method, they found that SiO$_2$, MgO, and FeO were the most common oxides, making up $\ge$80\% of the oxides determined from Hypatia stellar abundances, such that olivine and/or orthopyroxene likely dominate exoplanet mantle composition, with a smaller percentage of planets primarily made of magnesiow{\"u}stite or quartz. Their analysis also showed that exoplanets with exotic compositions, e.g., mantles made of albite, corundum, rutile, clinopyroxene, or garnet, may be incredibly rare, if they exist at all.

An observational test of the relationship between the refractory elements abundances in stars and their exoplanets was done by \citet{Bonsor21}. Because binary stars are conatal and should therefore have very similar compositions \citep[e.g.,][]{Hawkins20}, \citet{Bonsor21} examined the abundances of a K-dwarf star that is the wide binary companion to a polluted white dwarf. The polluted white dwarf had likely accreted a comet-like object from the system, making it possible to determine the composition of the comet-like body \citep{Xu17}. Since the comets, like planets, maintain a similar composition as the protoplanetary disk, they compared it to the chemical make-up of the companion K-dwarf star. They found that the refractory elements of the accreted material matched the abundances of the K-dwarf companion star to within errors (which they noted were to a higher accuracy than is typically achievable since the stars had similar stellar properties and were measured using the same technique, see their Figs. 2 and 3), thereby providing direct evidence that bodies within a planetary system and their host stars have the same refractory abundances. \citet{Guimond22} took this a step further and found that ``exoplanet mantle mineralogies are predicted in tandem with (i) measurements of stellar refractory element abundances and (ii) models calculating the equilibrium mineralogy for a given composition, pressure, and temperature."

\subsection{When The Star-Planet Compositional Link May Not Apply}

The most common means of estimating the compositions of small planets is through the relationship between their mass and radius. However, models based wholly on planetary mass and radius are replete with degeneracies. For example, issues are caused by the effects of volatiles on bulk densities as well as the uncertainties in the compositions of metal cores and rocky mantles. Regarding the latter, it was recently recognized that hydrogen, in addition to other light elements in iron-rich metal cores, can significantly change the mass-radius relationship for small planets \citep{Schlichting_Young_2022}. Also, the introduction of Si into the Fe-rich core, as a result of relatively reducing conditions at the core-mantle boundary, is accompanied by an increase in oxidized Fe in the rocky mantle \citep{Unterborn23}, altering the equations of state of both the core and mantle. Disambiguating the resulting changes in density and the potential presence of volatiles, as in the case of ocean worlds, is challenging at best \citep{Unterborn19}. A solution to break the mass-radius degeneracy is to use the abundances of rock-forming elements in stars as proxies for small planet compositions. However, this assumption sometimes breaks down, at which point other mechanisms or factors may need to be accounted for. 

\citet{Dorn19} studied the compositional relationship between super-Earths and their host stars. The authors began by recognizing that while the stellar disk is the remnant material from which planets form, the temperature of the disk controls when elements may condense, thereby creating compositional variations within different radial locations of the disk over time \citep[e.g.,][]{Lodders03}. While the alteration of the volatile elements are more substantial, the effect to refractory elements tends to be fairly minimal. The exception is when refractory elements partially condense in the inner-most part of the stellar disk, where temperatures can be $>$ 1200 K, thereby creating large compositional variations. It is within this regime that \citet{Dorn19} used stellar abundances from the Hypatia Catalog \citep[][see Section \ref{s.hypatia}]{Hinkel14} as proxies for the interior composition of orbiting planets and found that super-Earths, specifically HD~219134 b, 55 Cnc e, and WASP-47 e, could have formed with excesses in Al and Ca, but with cores that are depleted in Fe. These findings were followed up by \citet{Otegi20} who characterized the interiors of planets with masses $<$ 25 M$_{\oplus}$ and radii $<$ 3.5 R$_{\oplus}$, a middle ground between rocky and gas giant planets for which there are no equivalents within the Solar System. \citet{Otegi20} found that the interior planet models were not always better constrained when applying Fe/Si and Mg/Si molar ratios -- as determined from stellar abundances -- since it was dependent on the specific values of the stellar abundances and how they compared with the uncertainties in planetary mass and radius.

Super-Mercuries are also perplexing when it comes to the star-planet compositional link. Namely, Mercury is substantially iron-rich compared to the other small Solar System planets, such that its abundance of Si, Mg, and Fe does not align with the Sun, Earth, or Mars. It is thought that Mercury must have been subjected to a giant impact that stripped off all or part of its silicate mantle \citep{Benz08}. \citet{Bonomo19} discovered a similar exoplanet, Kepler-107 c, whose Fe-core fraction was so different that the density of the c-planet was twice that of its twin (Kepler-107 b). While many possible reasons were considered, the only mechanism that explained why only one of the twin planets was effected was a giant impact. \citet{Schulze21} expanded on the discovery of small, Fe-rich planets that did not have similar Fe/Mg abundances as the host star. They developed an agnostic planetary classifier that is able to determine probability density functions of the planet's core mass fraction calculated from the planet's mass and radius (CMF$_{\rho}$) and the core mass fraction determined using the relative abundances of Mg, Si, and Fe from the host star (CMF$_{star}$). They applied their methodology to 11 exoplanets, using abundances from the Hypatia Catalog \citep[][see Section \ref{s.hypatia}]{Hinkel14}, and found that the inferred interior properties of 9 of the exoplanets were indistinguishable from the host star. However, in two cases, the planets had densities that were notably different than predicted from the rock-building elements within the host star, such that Kepler 107 c was overly dense (likely super-Mercury) and 55 Cnc e was a low-density small planet (likely a mini-Neptune). Similar work was done by \citep{Unterborn23}, who examined planets around 7 stars using Hypatia Catalog (see Section \ref{s.hypatia}) abundances, where they found that 5 planets had CMF$_{\rho}$ $\approx$ CMF$_{star}$, to within respective errors, while the remaining 2 planets were overly dense, on par with super-Mercuries. The relationship between the two probability distributions, which account for all mass, radius, and abundance uncertainties, make it clear whether a planet's composition differs from the stellar abundances, similar to a super-Mercury or a mini-Neptune, or if it directly reflects the stellar abundances and is a nominally rocky planet \citep{Unterborn23}.

\begin{figure*}
\begin{center}
\centerline{\includegraphics[height=6.2in]{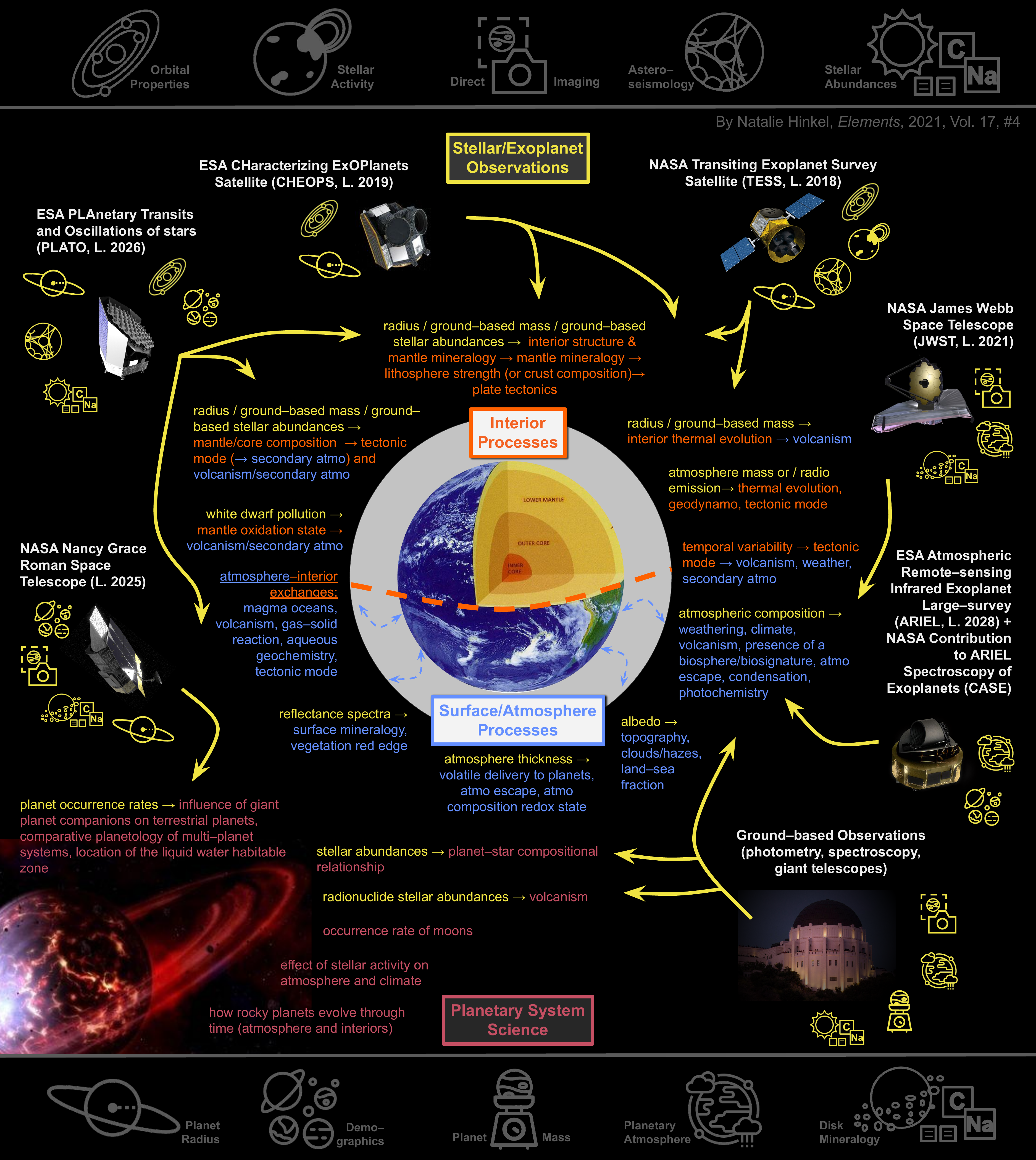}}
\end{center}
\vspace{-14mm}
\caption{An interdisciplinary map of the relationship between current/upcoming exoplanet data, i.e. from ground-based or space-based telescopes (yellow), and the way that it relates to our understanding of small planet interior processes (orange), surface and atmospheric processes (blue), or planetary system science (red). NASA and ESA missions (with respective launch dates, ``L.", provided), as well as ground-based observatories, are surrounded by icons (defined at the top and bottom of the figure) indicating the properties they are slated to measure. Large yellow arrows connect the mission observables to aspects of the planet and/or planetary system via smaller arrow flow downs. This figure created by Natalie Hinkel and featured in \citet[reproduced with permission from Elements Magazine]{Shorttle21}.}
\label{f.map}
\end{figure*}

\citet{Santos15} analyzed the question of planetary interior composition using stoichiometry, as opposed to interior models, where they balanced equations for the expected mass fractions of the dominant species (H, He, C, O, Mg, Si, and Fe) within a planet. This technique was used by \citet{Adibekyan21a, Adibekyan21b} who examined the make-up of 38 total small planets using stellar elemental abundances as determined from spectroscopic archival data (e.g, from HARPS, HIRES, HDS, etc.). They found that the planets did not reflect a 1-to-1 relationship with the host star abundances, but instead had an iron mass fraction that correlated 4-to-1 between the planet and star. They attribute this differentiation to planetary formation processes, such as a radial oxidation gradient within the protoplanetary disk or a combination of several mechanisms (such as Fe-enrichment in addition to a giant impact event). 

While the interior composition of small exoplanets is still a matter of intense study, it's clear that the incorporation of stellar abundances into planetary models is a benefit, especially for planet classification. Using the Solar system as a model, it seems that a planet is likely to reflect the abundances of rock-forming elements in the host star until it obviously doesn't (e.g., super-Mercuries and mini-Neptunes), in which case additional chemical or physical processes need to be considered. Testing the application of stellar abundances to smaller, possibly rocky planets suffers from a lack of data for small planets with both mass and radius measurements that orbit a star with refractory (Mg, Si, and Fe) element abundance determinations. Fortunately, there are a number of upcoming NASA and ESA missions that will be measuring these important stellar and planetary properties. Fig. \ref{f.map} shows how the observables from various upcoming missions (yellow), as well as ground-based observations, will provide key information for small planet interiors (orange), as well as surface/atmosphere processes (blue) and the planetary system science \citep[red, from][]{Shorttle21}. Fig. \ref{f.map} acts as an interdisciplinary map to connect the stellar and exoplanetary data we currently or will soon have with the ways in which they can be used to classify and better characterize small exoplanets, while also pointing out gaps in the data that have yet to be addressed.

\section{The Composition of Stars In the Solar Neighborhood} \label{s.neighborhood}

When trying to understand the composition of stars, it makes the most sense to turn first towards our closest star, the Sun. The proximity allows telescopes to achieve detailed measurements that aren't possible for any other star. In addition, it's possible to compare the elements measured within the Sun to the composition of other bodies within the Solar System, such as meteorites, asteroids, moons, etc.. For example, meteorites like carbonaceous chondrites were for the most part not subjected to physical or chemical fractionation processes within the stellar disk. With the exception of depleted volatile elements (H, He, C, N, and O), CI-chondrites (e.g., Ivuna-type carbonaceous chondrites) are believed to preserve the composition of many elements from when the Solar System was formed \citep{Lodders09}. It has proven useful to compare the photospheric elemental abundances of the Sun to the composition of meteorites (particularly CI-chondrites) as a way to verify the abundances and measurement techniques in both, especially since the Sun provides a complementary volatile element dataset \citep{Lodders19}.

However, the ability to compare abundances within the Sun to other Solar System bodies is only possible because of our location and proximity to those other bodies. We don't currently have the technical capability to directly measure the current interior composition of asteroids, moons, or even planets within exoplanet systems (although we can measure the make-up of rocky bodies accreted onto white dwarfs, see \S \ref{s.wd}). Therefore, to better understand the Sun, how it compares to nearby stars, as well as other stellar formation and/or evolutionary mechanisms, we analyze stars that are very similar to the Sun. These ``Solar twins" have stellar parameters that are within a narrow range of the Sun, e.g., effective temperature $\pm$ 100K, surface gravity $\pm$ 0.1-0.15 dex, and [Fe/H] $\pm$ 0.1 dex \citep{Nissen15, Nissen16, Bedell18}. It was long assumed that the Sun was a compositionally ``typical" star that had element abundances that were similar, if not the same, as other stars in the solar neighborhood. However, these high-resolution, differential, ultra-precise studies of the ``Solar twins" found that the Sun was distinctive in a number of ways. Namely, the Sun is depleted in the refractory elements or has a lower refractory-to-volatile element ratio compared with similar nearby stars \citep{Nissen15}, including $\sim$95\% of local ``Solar twins" \citep{Bedell18}. While the deficiency in refractory elements may be a signature of planet formation within the Solar System \citep{Melendez09, Ramirez09}, it is clear that the element abundance patterns within the Sun are unique. 

\begin{figure*}
\centering 
\includegraphics[height=5.0cm]{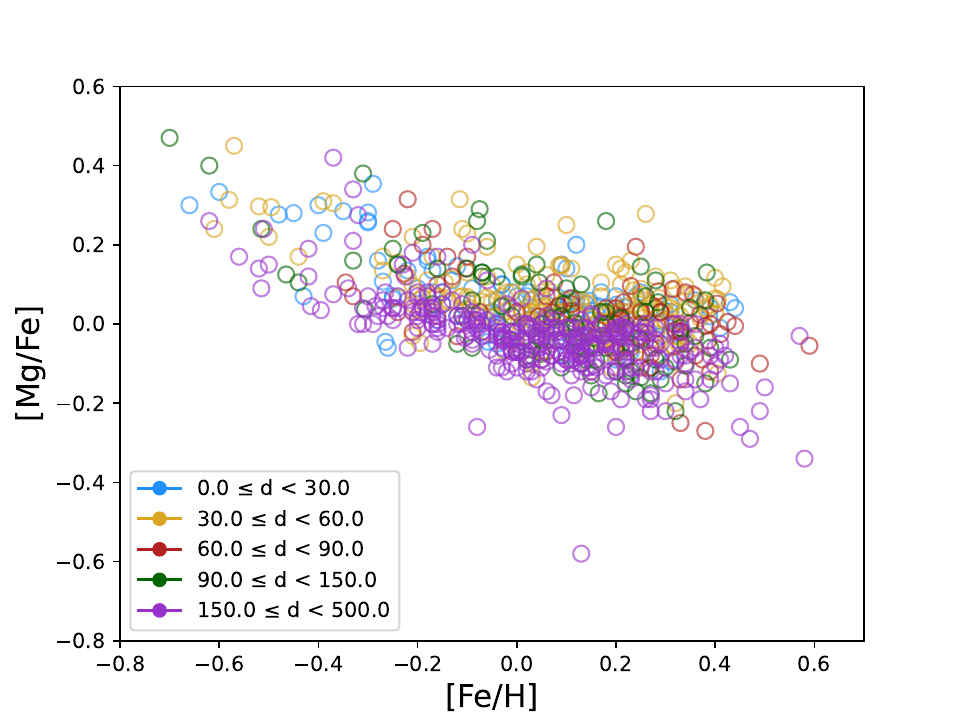}
\includegraphics[height=5.0cm]{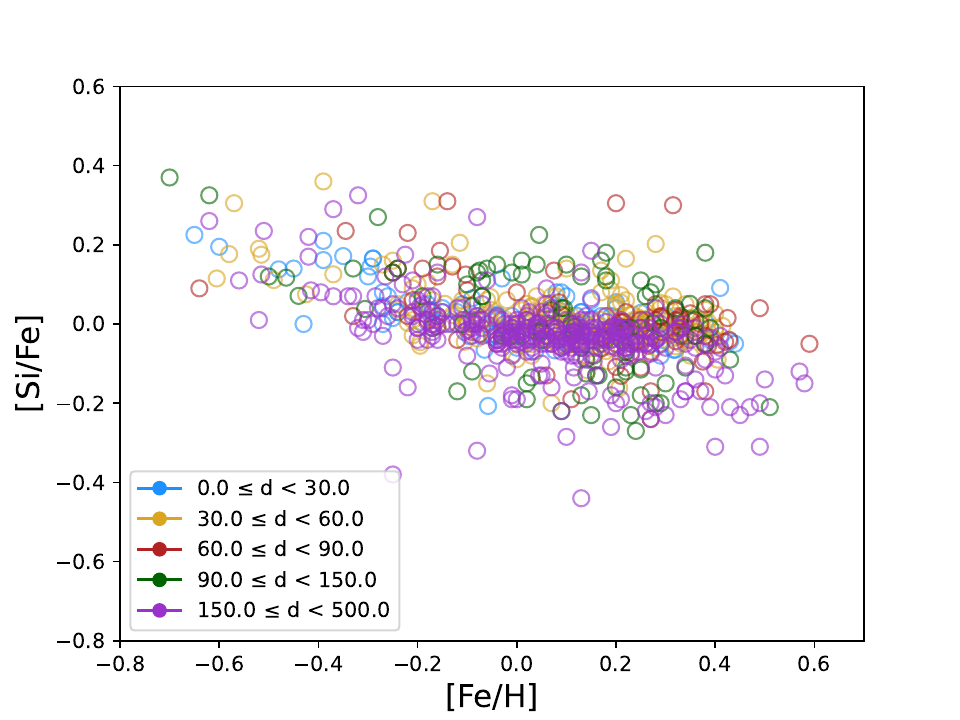}
\caption{Stellar abundances of exoplanet host stars found within the \HypCat \citep{Hinkel14} (see \S \ref{s.hypatia}) as compared to the Sun located at (0.0, 0.0) dex. All data points are color-coded to indicate the distance (d) in parsecs from the Earth. For a plot of stellar Fe/Mg vs Si/Mg using Hypatia Catalog data, we refer the reader to Fig. 3 in \citealt{Unterborn23}.}
\label{f.MgSi}
\end{figure*}

Therefore, to fully understand stellar formation, local chemical evolution, and the impact that planets may have on the host star's composition, it's important to look at larger patterns within the solar neighborhood to glean the more subtle patterns with stars, especially those that do and do not host planets. As an example, Fig. \ref{f.MgSi} provides the stellar abundances for all known exoplanet host stars within the \HypCat \citep{Hinkel14}, the largest database of nearby stellar abundances (\S \ref{s.hypatia}). We have specifically plotted [Mg/Fe] (left) and [Si/Fe] (right), both with respect to [Fe/H], since these three elements are the most important for building small planets. Given that the Sun is located at the center (0.0, 0.0) dex, it's clear that there is wide variety of stellar abundance compositions for exoplanet host stars compared to the Sun. Although, the chemical influence of giant planets vs small planets, multiplanetary systems, and the formation of super-Earths or mini-Neptunes -- all of which are being studied -- may have contributed to the range in distributions. Overall, though, Fig \ref{f.MgSi} shows that as the Fe content in the local Universe has increased (which tracks generally monotonically with time), the amount of Si and Mg have comparatively decreased -- namely that Mg and Si has not being created as quickly as Fe. It also shows that there is more scatter (perhaps due to more measurements) in the [Mg/Fe] abundances than in the [Si/Fe] abundances for planet-hosting stars.

\subsection{The Hypatia Catalog}\label{s.hypatia}
 
\begin{figure*}
\begin{center}
\centerline{\includegraphics[height=2.0in]{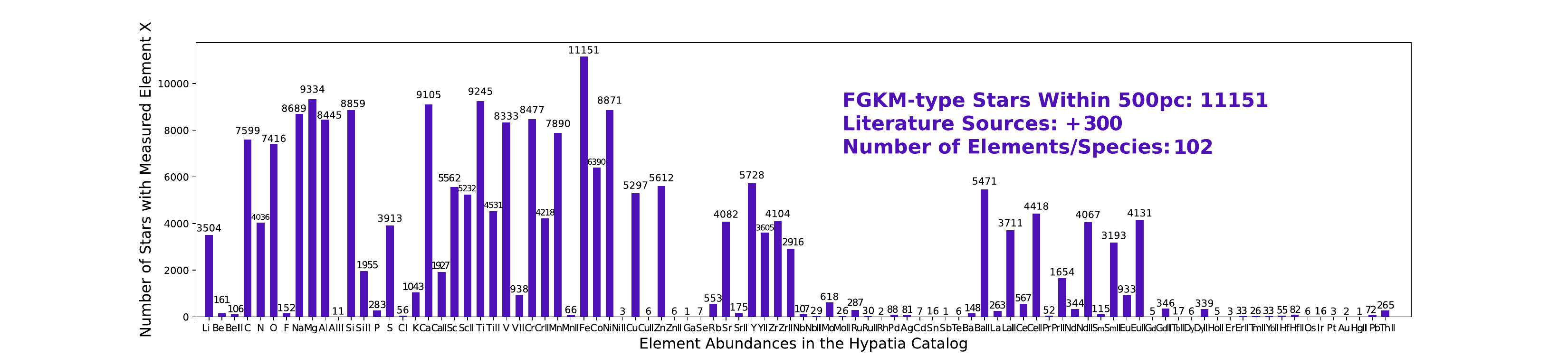}}
\end{center}
\caption{
Elements within the \textit{Hypatia Catalog} and the number of stars for which they've been measured. Of note are the elements that are often measured within stars versus those that are difficult to determine, but are important for planets and life such as N, F, P, Cl, K, and Mo.
}\label{f.hist}
\end{figure*}
	
The \HypCat (\url{www.hypatiacatalog.com}) is an amalgamate database that has been compiled from +300 literature sources to provide a comprehensive understanding of the abundances of $\sim$11,000 stars within the solar neighborhood \citep{Hinkel14}. Stellar abundances were included from any spectroscopic dataset that measured Fe and one other element for a main-sequence (FGKM-type) star within 500 pc, or any exoplanet host star regardless of distance, shown in Fig. \ref{f.hist}. Particular attention was paid to planet hosting stars (currently $\sim$1400 stars, per the NASA Exoplanet Archive), as well as multistellar systems (currently $\sim$2100 stars) since the occurrence of companions is likely to influence the composition of the stars. Additional stellar properties, e.g. position, magnitude, distance, velocity, temperature, and planet properties, e.g. mass, radius, eccentricity, are included within \Hyp to better understand the kinematic and chemical patterns present within nearby stars. For example, different galactic populations may be discerned through stellar kinematics and composition \citep{Freeman02}. Therefore, the likely population membership, namely thin-disk, thick-disk, or the halo of the Milky Way, was also determined for each of the stars, based on the conservative kinematic prescription per \citet{Bensby03}.

By design, the \HypCat is a multidimensional database that features the element measurements for each star as measured by every literature source, totaling $\sim$370,000 abundance measurements. However, as discussed in \S \ref{s.spectechniques}, each stellar abundance study uses different telescopes, stellar atmospheric models, spectral fitting techniques, line lists, solar normalizations, etc., which imparts systematic offsets between the abundance determinations. While most of these variations are inherent to the element measurements, the solar scale is a product of the finalized abundance-dex-notation \citep{Hinkel22} that can be altered after-the-fact. Therefore, in order to be certain that all abundances are on the same solar baseline, all datasets within \Hyp have the original solar normalization removed in lieu of a standardized solar scale. As reported in \citep{Hinkel14}, there was an average 0.06 dex (0.04 dex median) variation between the abundances before and after the solar re-normalization. This variation is larger than most errors associated with the elemental abundances, indicating that the choice of solar normalization has a strong impact on the final abundance determination. While the \HypCat has currently encountered $\sim$75 solar normalizations, only $\sim$10 originated from studies that were specifically dedicated to solar abundance determination -- see \citet{Hinkel22} for a brief overview and comparison of the most popular dedicated solar studies.

Overall, the \HypCat allows researchers to view nearly every measurement of an element within a star that have ever been published. The benefit is that the field is no longer dependent on singular benchmark datasets \citep[e.g.,][]{Edvardsson93, Thevenin1999, Nissen10, Bensby14} in order to validate methodology and/or compare new results. This is exceptionally important given the subtle variations between stellar abundance methodologies that may not yield apples-to-apples comparisons. In addition, by having all of the data in one centralized repository, scientists who are not experts in stellar spectroscopy, especially those outside of the astronomy field altogether, no longer have to go hunting for the best, latest, or most commonly used dataset for their research application, thereby reducing siloing while encouraging interdisciplinary science.

\begin{figure*}[ht!]
\begin{center}
\centerline{\includegraphics[height=4.0in]{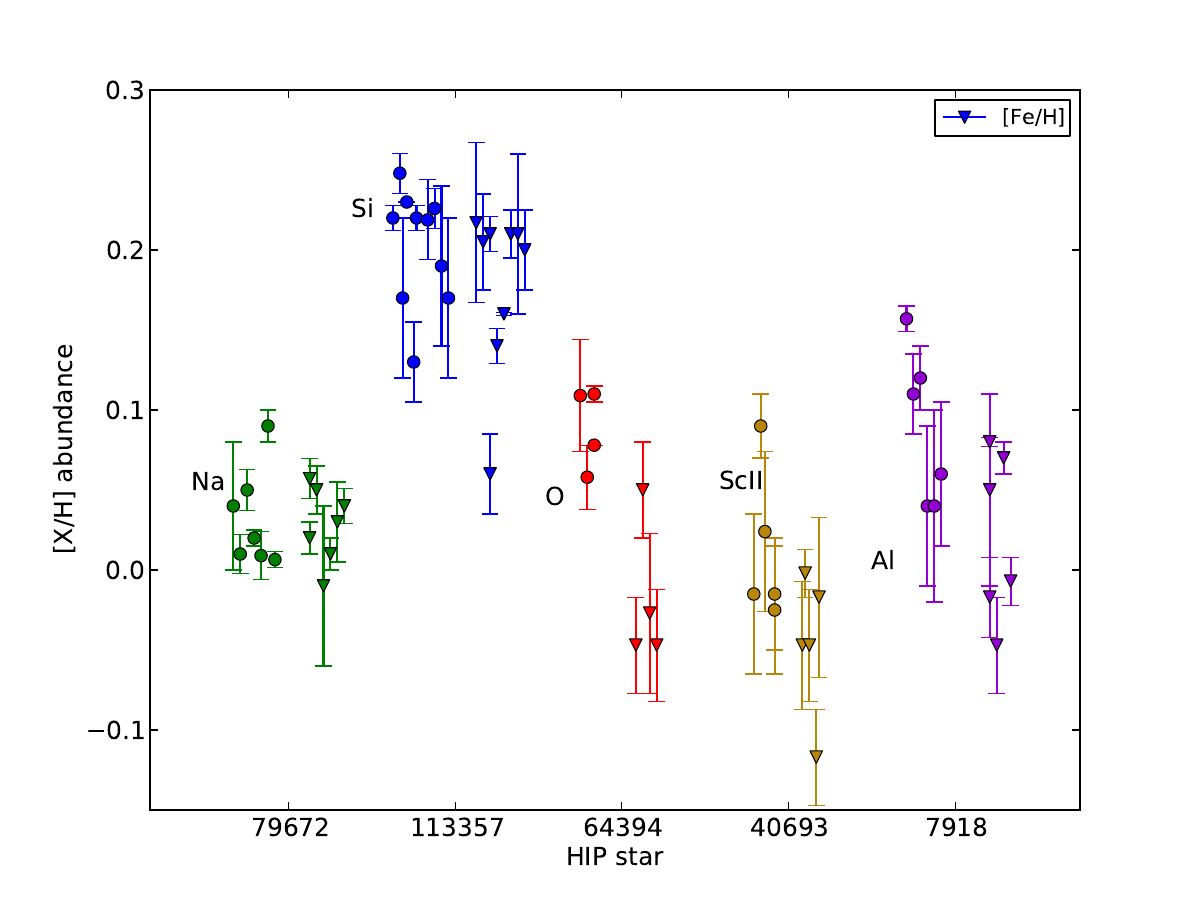}}
\end{center}
\vspace{-15mm}
\caption{Reproduction of Fig. 3 (top) from \citet{Hinkel14} showing the variation between datasets when measuring the same elements within the same star, incorporating the individually reported error. The circles indicate (from left to right): [Na/H] (green), [Si/H] (blue), [O/H] (red), [Sc II/H] (yellow), and [Al/H] (purple). The [Fe/H] measurements associated with the [X/H] determinations are provided as triangles, also with individual error. The five unique stars (x-axis) were chosen because they were often measured, i.e. by multiple literature sources. Reproduced by permission of the AAS.
}\label{f.spread}
\end{figure*}

The consequence of all the compiled data within the \HypCat is that variations between datasets, specifically those studying the same element within the same star, are immediately apparent. Fig. \ref{f.spread} illustrates how multiple literature sources did not measure consistent elemental abundances for six elements (Na, Si, O, Sc II -- or singly ionized scandium, and Al in circles and their corresponding Fe measurements in triangles) to within their individually reported error bars \citep{Hinkel14}. While this issue was later addressed by multiple teams (see \S \ref{s.spectechniques}), the range between all of the abundance measurements per star is an extremely useful indicator for how well each element is truly known within a star. Therefore, the range in measurements, i.e. the maximum minus the minimum, is called the \textit{spread} within \Hyp and is used to represent the error or uncertainty in that element. Because stellar abundance error may be reported on a star-by-star basis or for each element as a whole, the average reported uncertainty is used, or ``representative" error, is used in those instances where only one literature source measured the abundances of a star.

Compiling multiple datasets into one large repository for the \HypCat inherently creates a heterogeneous, as opposed to a homogeneous, database. Namely, the caliber of abundances and total number of elements available is prone to vary for each star. However, the elemental abundances often originate from specialized studies using high-resolution, ground-based data. So while the individual differences between literature sources may require more attention when analyzing a particular stellar or planetary system, the investigation ultimately results in a more comprehensive understanding of the true nature of the system since it has been analyzed with multiple techniques. Bringing together such a large number of datasets also enables the most comprehensive suite of element abundances available, or 2 or 3 times the number of elements available through large surveys (e.g., APOGEE, GALAH, RAVE, etc), as shown in Fig. \ref{f.hist}. Per a study by \citet{Spaargaren22}, they compared the abundances from the \HypCat to GALAH for stars within 200 pc and found no systematic differences in the abundances of both catalogs. It is therefore the combination of the \HypCat's depth (number of elements) and breadth (number of stars) that render it such a powerful tool. 

\subsection{Large Stellar Abundance Surveys}

Homogeneous datasets feature measurements that originate from the same telescope, reduction pipeline, and overall analysis. Large spectroscopic surveys are able to specifically choose particular populations of stars, e.g., focusing on certain stellar types, a volume-limited sample, a magnitude-limited sample, etc.. They observe thousands of stars whose abundances can be easily compared with one another without the need to correct for systematic offsets, which makes it easier for observing large-scale patterns across the galaxy. Here we describe some of the current or upcoming large-scale surveys (in order of lowest to highest spectral resolution):

\vspace{2mm}
\noindent
\underline{RAVE}: The RAdial Velocity Experiment (RAVE) survey is a magnitude-limited spectroscopic survey that uses the 1.2 m Schmidt Telescope at the Anglo-Australian Observatory \citep{Steinmetz06}. The telescope observes from 8410-8795 \AA\, with a resolution of R$\sim$7500 and S/N$\sim$ 50/pixel. As a result, they have observed 7 elements (Mg, Al, Si, Ca, Ti, Fe, and Ni) with an error of $\sim$0.2 dex for $\sim$300,000 stars in the Milky Way.

\vspace{2mm}
\noindent
\underline{WEAVE}: The WHT Enhanced Area Velocity Explorer (WEAVE) survey will soon come online and will use a dedicated spectrograph on the 4.2-m William Herschel Telescope (WHT) to obtain follow-up observations necessary for previous surveys \citep{Dalton12}. WEAVE will measure abundances for $\sim$1.5 million bright-field and open-cluster stars from 3660-9590 \AA\ in its high-resolution (R $\sim$ 20,000) mode. While specific elements have not been named, WEAVE will cover elements from a variety of nucleosynthetic channels, such as volatiles, $\alpha$-elements, Fe-peak, as well as neutron capture (s- and r-process) elements, to an anticipated errors of 0.1-0.4 dex \citep{Jin22}. 

\vspace{2mm}
\noindent
\underline{APOGEE}: The Apache Point Observatory Galaxy Evolution Experiment (APOGEE) is an infrared spectroscopic survey (1.514 to 1.696 $\mu$m) that uses the 2.5-m Sloan Telescope and has an R$\sim$22,500 and S/N $>$ 100 \citep{Majewski17}. The survey measures 20 elements (C, N, O, Na, Mg, Al, Si, P, Si, K, Ca, Ti, V, Cr, Mn, Fe, Co, Ni, and Rb) with an error of $\lesssim$ 0.1 dex within $\sim$300,000 giant stars in the galaxy.

\vspace{2mm}
\noindent
\underline{GALAH}: GALactic Archeology with HERMES (GALAH) survey is based on the 3.9-m Anglo-Australian Telescope and uses the High-Efficiency and Resolution Multi-Element Spectrograph \citep[or HERMES per][]{Buder21}. While they have currently observed $\sim$350,000 stars, they plan to observe $\sim$1 million stars with a resolution of R $\sim$ 28,000 at a wavelength range of 4713-7887 \AA\,. Determine $\sim$30 element abundances (Li, C, O, Mg, Si, Ca, Ti, Na, Al, K, Sc, V, Cr, Mn, Fe, Co, Ni, Cu, Zn, Rb, Sr, Y, Zr, Ba, La, Ru, Ce, Nd, and Eu), to an accuracy of $\sim$0.05-0.20 dex.

\vspace{2mm}
\noindent
\underline{Gaia-ESO}: The 8-m Very Large Telescope was the home of the Gaia-ESO survey, which uses the Fibre Large Array Multi-Element Spectrograph \citep[or FLAMES per][]{Gilmore12}. FLAMES consists of the Ultraviolet and Visual Echelle Spectrograph (UVES) -- which observes at 3000-11000 \AA\, with a high resolution of R $\sim$ 47,000, in addition to the Giraffe Spectrograph which has a medium resolution R $\sim$ 20,000. Gaia-ESO was able to observe $\sim$115,000 stars within many parts of the Milky Way and observed $\sim$ 30 elements (Li, C, N, O, Na, Mg, Al, Si, S, Ca, Sc, Ti, V, Cr, Mn, Co, Ni, Cu, Zn, Y, Zr, Mo, Ba, La, Ce, Pr, Nd and Eu) with errors typically between 0.1-0.2 dex.

\subsection{Spectroscopic Techniques}\label{s.spectechniques}
Determining stellar elemental abundances is a combination of theoretical physics (e.g., stellar atmospheric models, line formation, atomic/molecular transitions), experimental laboratory results (e.g., atomic parameters, line broadening, radiative transition probabilities), and observational astronomy (application to observed stellar spectra); we refer the reader to a more in-depth discussion by \citet[][and references therein]{AllendePrieto16, Jofre19}. Stellar atmospheric models emulate the processes and exchanges that occur in the top layers of the star, defining the ways that the stellar gas interacts with radiating photons from deep within the stellar interior. Photons emit, reabsorb, and/or scatter light that later encounters atoms or molecules which then absorb the light at specific wavelengths, per their different discrete energy levels, creating absorption lines in the stellar spectra. The width and depth of the absorption line, when accounting for stellar temperature and pressure, is directly correlated to the number of absorbing atoms/molecules. From here, using either a synthetic spectra or the equivalent width technique, it is possible to determine the specific number of atoms/molecules within a star and determine its overall abundance. 

The determination of spectroscopic, as opposed to photometric, stellar abundances gained traction in the 1970s, courtesy in no small part to \citet{Sneden73} and \citet{Gustafsson75}. They produced some of the first radiative transfer and atmospheric models, respectively -- both of which have been maintained and are still popular tools today. Since that time, though, a variety of model atmospheres and radiative transfer codes, in addition to grids of synthetic stellar spectra and software for automatic/batch abundance determinations, have been created within the community \citep[see Table 1 in ][for a thorough list of regularly updated, publicly available tools]{Jofre19}. There are also a large number of atomic and molecular line lists, which may have been compiled from literature, theoretically calculated, or empirically derived within a laboratory. These line lists contain the wavelength, transition probabilities, and atomic properties necessary for assessing the spectral absorption features for abundance determination, however, they are also the source of large uncertainties and inconsistencies. For example, most line lists are focused on atoms within the optical wavelength range whose lines are common in solar-like stars, and even then only $\sim$50\% of these lines have robust laboratory transition probabilities with errors $\lesssim$ 10\% \citep{Jofre19}. Line lists for cooler stars -- which are often dominated by large molecular bands within their spectra, as well as spectral observations in the ultraviolet (UV) or infrared (IR), are often incomplete and/or piecemeal. This is to say nothing of the influences that different telescopes and their spectrographs may have on the measured element abundances. Namely, the resolution of the spectrograph will determine whether specific spectral lines can be resolved as well as the presence of line asymmetries, blends, etc., such that R $\gtrsim$ 50,000 in the optical and R $\gtrsim$ 20,000 in the IR are considered high-resolution. The signal-to-noise (S/N) of the instrument dictates the precision to which the stellar abundances can be calculated, where high S/N is typically defined as S/N $>$ 100. There are additional issues to consider, such as time dependent phenomena like stellar pulsations and starspots, data reduction, telluric interference from the Earth's atmosphere, flux calibration, and flux calibration -- which are explained in more detail by \citep{Jofre19}. 

Overall, there are a huge number of choices and variations that can occur within stellar abundance analyses, creating discrepancies between the different methodologies -- as was exemplified in Fig. \ref{f.spread}. We therefore refer the reader to \citet{Smiljanic14}, \citet{Hinkel16}, and \citet{Jofre17} who analyzed multiple stellar abundance techniques using a variety of methods from within the stellar abundance community to understand their impact on abundance determinations.

\section{Polluted White Dwarfs}\label{s.wd}
White dwarfs (WDs) polluted by accretion of asteroid-like rocky bodies offer an alternative means of estimating the compositions of rocky bodies in the galaxy. This approach has the advantage that the compositions of rocky bodies themselves are interrogated, rather than stellar proxies, but at the cost of less precision compared with many stellar observations. Typical WDs are the result of A- or F-type stars that are no longer able to fuse H into He, as part of their main sequence lifetime, and eventually lose much of their mass as they evolve through the AGB phase to form WDs. These WDs have masses typically similar to $0.6$ M$_{\odot}$, but radii comparable to Earth. The cores are made primarily of oxygen and carbon but the near-surface environs are composed of residual He and sometimes H. The resulting immense gravitational field causes ``metals" (how astronomers refer to elements heavier than He) to sink rapidly from the surface into the interior of the WD. Therefore, spectral lines of rock-forming elements coming from the surfaces of these stars provide evidence for the accretion of rocky debris; any heavy elements seen in metal spectral lines are from exogenous rock-forming elements. Decades of work has shown that the elemental ratios obtained by observations of WDs with heavy elements are best explained as representing accretion of rocky bodies often broadly similar in mass to asteroids from our Solar System \citep[e.g.,][]{Zuckerman07,Klein2010,Melis2011,Farihi2011,Zuckerman2011, Jura2012,Gansicke2012,jura2013al26,Jura_Young_2014,Xu17,Swan_2019,Doyle2019,Bonsor2020,Trierweiler2022}. Direct evidence for pollution of WDs by debris from rocky bodies has been found from transit studies \citep{Vanderburg2015, Manser_2019}. 

Surface temperatures of WDs exhibiting observable pollution vary from approximately $20,000-5000$ K, depending on their cooling ages of $10^7$ to $10^9$ years, respectively \citep{Zuckerman_2017}. They are categorized according to their spectral types. In the broadest sense, two primary categories are defined. Those with spectra dominated by He lines are referred to as ``DB" stars (where D refers to ``degenerate"). Those with spectra dominated by H lines are referred to as ``DA" stars. In addition, where metals \citep[Z, see][]{Hinkel22} lines are especially visible, a ``Z" can be added. For example, identifying DBZ, or even DBAZ where He and H lines are present, with the order representing the relative strengths (equivalent widths) of the spectral lines. The data comprise values for $\log{(Z/X)}$ where $Z$ represents the atomic abundances of metals, including important rock-forming elements like Si, Mg, Ca, Al, Fe, Ti, Cr, N and O, and $X$ is the atomic abundance of either H (DA WDs) or He (DB WDs). Relative abundances are obtained from equivalent widths of lines from fits of the spectra with models for the gravitational accelerations ($g$) and temperatures at the surfaces of the WDs. Uncertainties in the logs of ratios are usually reported as symmetrical errors (e.g., $\log{Z/X} = -4.5 \pm 0.2$). The uncertainties include those arising from the fits of the models to the spectra. 

Accretion of debris captured by a WD occurs in three phases \citep{Koester2009} that can be described by calculations that consider the competing effects of the time-dependent accretion from the debris disk surrounding the WD and settling through the atmosphere of the WD \citep{Jura2009,Doyle20}. The functions relating the relative elemental abundances in the polluted WD to those of the accreting parent body differ from phase to phase. Initially, accretion leads to a build up in the outer layer of the WD on timescales shorter than characteristic settling times through these layers. The build-up phase transitions to a steady state between accretion and settling, followed by a dominance of settling as accretion wanes. 
In this context, the distinction between DA and DB WDs is an important one. Heavy elements in DB WDs settle on timescales of $10^3$ to $10^6$ years, depending upon age and thus temperature, whereas metals in DA stars settle in days \citep{Koester2009}. In lieu of more detailed models for the time evolution of accretion, element concentration ratios for the parent rocky bodies can be derived from those observed in the WDs using a generic equation for the time-dependent addition of element Z resulting from addition to the convective layers or photospheres of the WDs:

\begin{equation}
\frac{{d{M_Z}}}{{dt}} = {\dot M_Z} - \frac{{{M_Z}}}{{{\tau _Z}}},
\label{eq:dmzdt}
\end{equation}
where $M_Z$ is the mass of element $Z$ in the WD atmosphere, $\dot M_Z$ is the accretion rate of the element onto the star, $t$ is the elapsed time for accretion, and $\tau_Z$ is the e-folding time for settling out of the convective layer or photosphere for element Z. This equation has the general solution

\begin{equation}
{M_Z} = c{e^{ - t/{\tau _Z}}} + {e^{ - t/{\tau _Z}}}\int {{e^{t/{\tau _Z}}}} {\dot M_Z}(t)\,dt,
\label{eq:wd_solution}
\end{equation}
where $c$ is an integration constant that is zero where the mass of $Z$ at time zero is zero. For the He-rich DB WDs with temperatures $< 17000$K, $\tau_z$ are long compared with accretion times $t$, and we have

\begin{equation}
{M_Z} = \int {{{\dot M}_Z}(t)} \,dt.
\label{eq:DBs}
\end{equation}
With long settling times of order $10^6$ years, the prospect for a buildup of the polluting rock-forming elements during accretion in the atmosphere arises as the system approaches a steady state between the rate of accretion and settling. Where build up is occurring, the ratio of any two elements observed in the WD atmosphere, $Z_1$ and $Z_2$, faithfully reflects the ratio for the accreting parent body because then

\begin{equation}
\frac{{{M_{{Z_2}}}}}{{{M_{{Z_1}}}}} = \frac{{{{\dot M}_{{Z_2}}}}}{{{{\dot M}_{{Z_1}}}}}
\label{eq:buildup}
\end{equation}
so the ratio of accretion rates corresponds to the element ratio in the parent body of the accreted rocky material. Steady state is a relatively brief interval during the accretion episode for DB WDs \citep{Doyle20}. Conversely, for DA WDs, the settling timescales are much shorter than the duration of the accretion events, and the polluting material is only visible if a steady state between accretion and settling is achieved. In these cases, $e^{({-t/\tau_z})}$ approaches zero, and we have 

\begin{equation}
\begin{split}
{M_Z} & = {e^{ - t/{\tau _Z}}}\int {{e^{t/{\tau _Z}}}} {{\dot M}_Z}(t)\,dt \\
 & = {{\dot M}_Z}{\tau _Z}\left( {1 - {e^{ - t/{\tau _Z}}}} \right) \\
 & = {{\dot M}_Z}{\tau _Z}.
\end{split}
\label{eq:steadystate}
\end{equation}
Since the element ratios in the parent accreting body determine the accretion rate ratio, Equation \ref{eq:steadystate} shows that parent body element ratios at steady state are related to those in the WD by the inverse ratio of characteristic settling times: 

\begin{equation}
{\left( {\frac{{{M_{{Z_2}}}}}{{{M_{{Z_1}}}}}} \right)_{\rm PB}} =
{\left( {\frac{{{M_{{Z_2}}}}}{{{M_{{Z_1}}}}}} \right)_{\rm WD}}\frac{{{\tau _{{Z_1}}}}}{{{\tau _{{Z_2}}}}}.
\end{equation}
In practice, settling times are obtained from estimates from theory that in turn rely on models for the temperature and gravity of the WD \citep{Koester2009}. The settling time corrections are usually on the order of a factor of 2 or less.

Detailed geochemical studies of extrasolar rocks have been afforded by polluted WDs. Results show that rocks beyond the Solar System have oxidation states similar to chondrite meteorites, for example \citep{Doyle19}. Recently, substantial excesses in the light elements Be and Li have been discovered in polluted WDs \citep{Klein2021,Kaiser2020,hollands2021}. The excess Be has been interpreted as being the result of spallation of water ices in the magnetic field surrounding a giant planet \citep{Doyle2021} while the Li has been variously interpreted as being either evidence for accretion of highly differentiated crust \citep{hollands2021} or a remnant of Big-Bang nucleosynthesis \citep{Kaiser2020}. Correlations between element ratios indicate igneous differentiation in a number of polluted WDs \citep{Jura_Young_2014}. Iron core-rich material accreted by WDs attests to complete silicate-metal differentiation in some cases \citep{Harrison2018,Bonsor2020}, while a primitive Kuiper-Belt like body is evidenced in one polluted WD \citep{Xu2017}. 

While such studies have revealed some surprisingly detailed aspects of the geochemistry of extrasolar rocks polluting WDs, as an ensemble, the rocks polluting WDs appear to be very similar to those in our Solar System, including in their relative abundances of the major rock-forming elements \citep{Jura_Young_2014}, the general depletion of carbon relative to solar abundances \citep{jura2013review}, and the occasional presence of modest amounts of water \citep{Farihi2013}. Indeed, most rocks are indistinguishable from either chondrites, or bulk silicate Earth. As an illustration, here we show the results of a comparison between the 31 polluted WDs that at the time of this writing comprise those WDs with sufficient pollution to allow characterization of the bulk rock composition of the parent polluting rocks to chondrite meteorites. 

CI chondrites are used as the benchmark for archetypal, primitive rock compositions in the Solar System. Here we compare rock compositions obtained from the 31 polluted WDs to CI chondrite using the reduced $\chi^2$ values for correlations between the WD pollutants and CI, such that
\begin{equation}\label{eq:chi2reduced}
\begin{split}
 \chi^2_\nu= \frac{\chi^2}{\nu}= \frac{1}{n} \sum_i \frac{(C_{i,{\rm WD}} - C_{i,{\rm CI}})^2}{\sigma_i^2}, 
\end{split}
\end{equation}
\noindent where $n$ is the number of samples of data used (i.e. the number of elements considered), $C_{i,{\rm WD}}$ is the concentration of element $i$ derived from the WD, $C_{i,{\rm CI}}$ is the concentration of that element in CI chondrite, and $\sigma_i^2$ is the variance associated with the measurement uncertainty. Measurement uncertainties for the WDs are propagated using a Monte Carlo approach. We use ratios of the elements Si, Fe, Al, Ca, Ni, and Cr and Ti, where available, to Mg, as relative concentrations. Asymmetric errors in element ratios arising from reported symmetric errors in logs of the ratios $Z/X$, are accounted for using \citep{Barlow2003}
\begin{equation}
 \chi^2 = \left(\frac{\delta}{\sigma}\right)^2 \left(1 - 2A \frac{\delta}{\sigma} + 5A^2 \left(\frac{\delta}{\sigma}\right)^2 \right),
\end{equation}
\noindent where $\delta$ is the difference between the observed and expected element ratios, $\sigma$ is the average of the upper and lower errors (16.5 and 83.5 percentiles), and $A = (\sigma_+ - \sigma_-)/(\sigma_+ + \sigma_-)$ is the asymmetry factor. In addition, because the number of elements is small, uncertainties in the $\chi^2_\nu$ values themselves must be considered. The error in $\chi^2_\nu$ can be approximated as $\sigma = \sqrt{2/n}$ \citep{Andrae2010}, where again $n$ is the number of data points for a given star's composition. With this approach, one can define a critical $\chi^2_\nu$ value: $\chi^2_{\nu, \rm crit} = \chi^2_\nu (\alpha = 0.05) + 2 \sqrt{2/n}$, allowing for the $2\sigma$ error in $\chi^2_\nu$. These constraints give critical $\chi^2_\nu$ values of $\sim 3-5$, depending on the number of elements observed for WD. If element ratios for a WD yield $\chi^2_\nu$ values $\lesssim 3-5$, the data are taken as evidence for chondritic rocky parent bodies or planets. The value for $\alpha$ defines the probability of randomly obtaining a $\chi^2$ value greater than that calculated for the observed abundances (e.g.\, the probability of incorrectly rejecting the null hypothesis, $H_0$, that the rocks are not chondritic). With $\alpha = 0.05$, $H_{\rm a}=1-\alpha$ is the probability that the correspondence with chondrite is not due to random chance. 

\begin{figure}
\vspace{-17mm}
\begin{center}
\centerline{\includegraphics[width=5.0in]{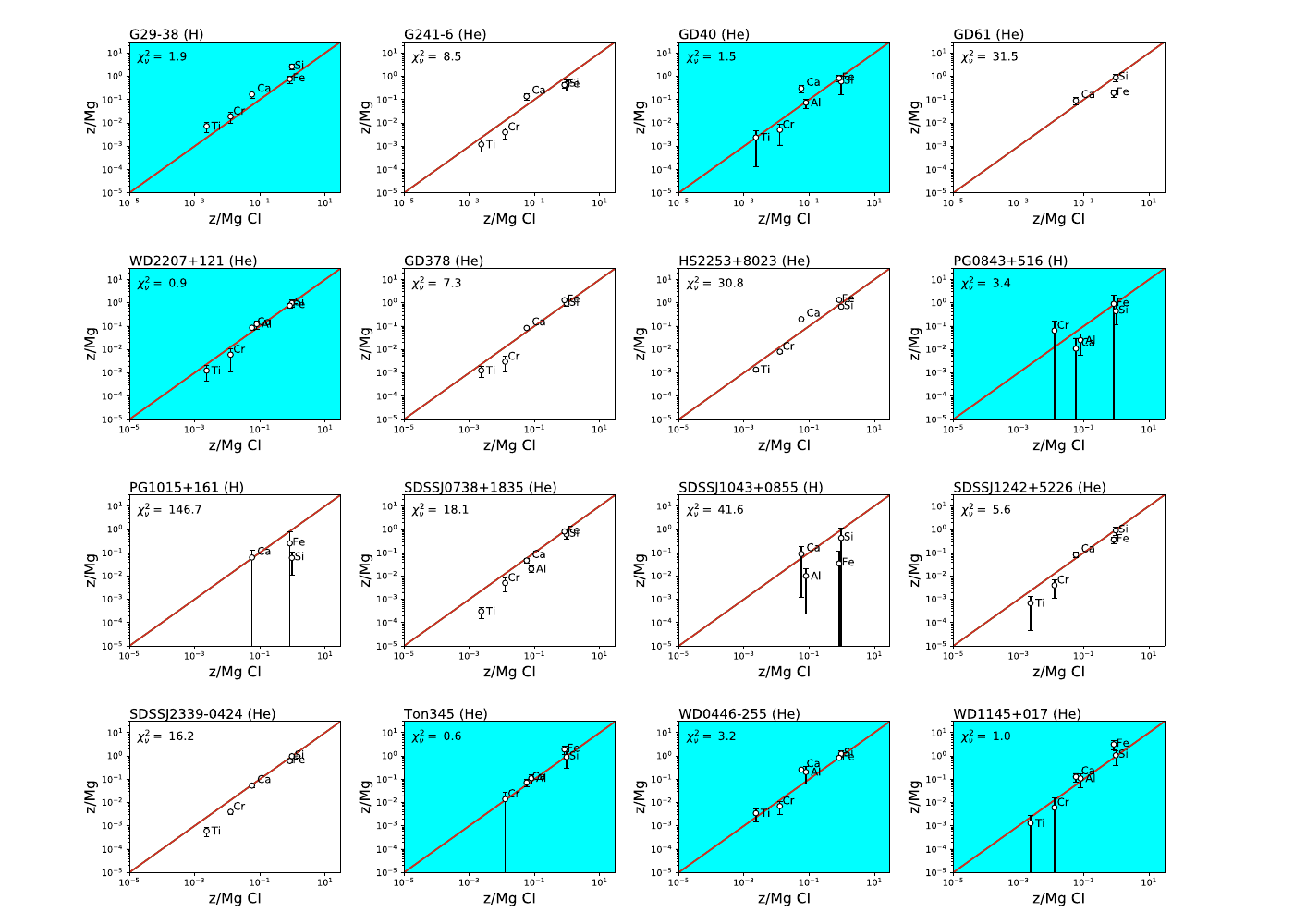}}
\end{center}
\vspace{-14mm}
\caption{First of two tranches of results of correlation analysis between polluted WDs and CI chondrites. Blue shaded panels are those WDs that yield rock compositions indistinguishable from CI chondrite. White panels have reduced chi-square values above the critical value described in the text. Data sources are as follows: 
G29-38\citep{Xu2014}, 
G241-6 \citep{Jura2012},
GD40 \citep{Jura2012},
GD61 \citep{Farihi2013}, 
WD2207+121 \citep{Xu2019},
GD378 \citep{Klein2021},
HS2253+8023 \citep{Klein2011},
PG0843+516 \citep{Gansicke2012},
PG1015+161 \citep{Gansicke2012},
SDSSJ0738+1835 \citep{Dufour2012},
SDSSJ1043+0855 \citep{Melis2017},
SDSSJ1242+5226 \citep{Raddi2015},
SDSSJ2339-0424 \citep{Klein2021},
Ton345 \citep{Wilson2015}, 
WD0446-255 \citep{Swan2019}, and 
WD1145+017 \citep{Fortin-Archambault2020}.
}
\label{f.wds1}
\end{figure}

\begin{figure}
\vspace{-17mm}
\begin{center}
\centerline{\includegraphics[width=5.0in]{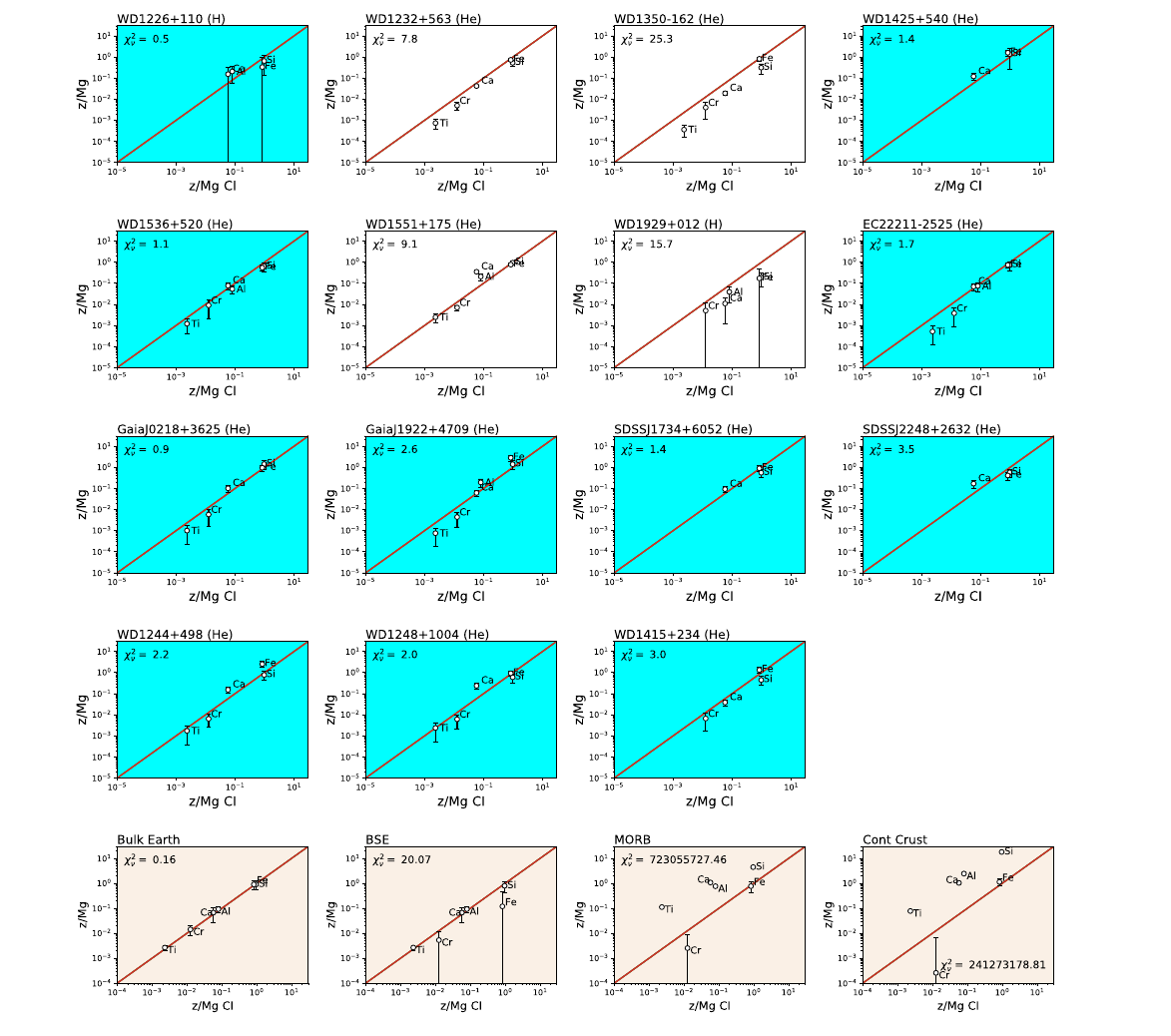}}
\end{center}
\vspace{-14mm}
\caption{Second of two tranches of results of correlation analysis between polluted WDs and CI chondrites. Blue shaded panels are those WDs that yield rock compositions indistinguishable from CI chondrite. White panels have reduced chi-square values above the critical value described in the text. Analogous correlations between CI chondrite and bulk Earth, bulk silicate Earth (BSE), mid-ocean ridge basalts (MORB), and continental crust are shown for comparison in the beige panels. Data sources are as follows: 
WD1226+110 \citep{Gansicke2012},
WD1232+563 \citep{Xu2019},
WD1350-162 \citep{Swan2019},
WD1425+540 \citep{Xu2017},
WD1536+520 \citep{Farihi2016},
WD1551+175 \citep{Xu2019},
WD1929+012 \citep{Gansicke2012},
EC22211-2525 (Doyle et al. in prep),
GaiaJ0218+3625 (Doyle et al. in prep).
GaiaJ1922+4709 (Doyle et al. in prep),
SDSSJ1734+6052 (Doyle et al. in prep),
SDSSJ2248+2632 (Doyle et al. in prep),
WD1244+498 (Doyle et al. in prep),
WD1248+1004 (Doyle et al. in prep),
WD1415+234 (Doyle et al. in prep), 
bulk silicate Earth (BSE) \citep{McDonough2003},
mid-ocean ridge basalts (MORB), and 
continental crust \citep{Rudnick2003}.}
\label{f.wds2}
\end{figure}

Figures \ref{f.wds1} and \ref{f.wds2} show the results of this comparison. Eighteen of the 31 WDS for which we have sufficient data are statistically indistinguishable from CI chondrite. It is important to note that when constrained by the  uncertainties in the WD element ratios, as in Figure \ref{f.wds2}, bulk Earth \citep{McDonough2003} and CI chondrite are also indistinguishable (Figures \ref{f.wds1} and \ref{f.wds2}), while bulk silicate Earth (BSE) is marginally distinguished from CI chondrite. Terrestrial crust, either oceanic crust represented by mid-ocean ridge basalts (MORB), or continental crust, are readily distinguished from CI chondrite using this method. For the WDs that are not statistically identical to CI chondrite (or bulk Earth), the $\chi^2_\nu$ values are usually within a factor of 2 to 5 of the critical values, as opposed to factors of many orders of magnitude obtained from MORB or continental crust. This suggests that these WDs are only marginally distinct from CI chondrite. There is no clear evidence for crust in these data, and similarly no clear evidence for rock compositions that might be considered to be unusual in the context of Solar System bodies (Figures \ref{f.wds1} and \ref{f.wds2}). This conclusion is contrary to some work that has suggested that the polluted WDs exhibit evidence for unusual rock types relative to solar \citep[e.g.,][]{Putirka_Xu_2021}. The difference in conclusions is likely due to variations in propagation of uncertainties.

For more than a century, geochemists have represented rock chemistry as ``normative mineralogies" in which elemental concentrations are converted to volumetric fractions of fictive minerals \citep{Cross1902}. Normative mineralogies have the benefit of relating bulk rock chemistry to tangible minerals used to classify rocks and it's an important method for discussing chemistry from a mineralogical perspective. \cite{Putirka_Xu_2021} used this approach for rock compositions derived from polluted white dwarfs and concluded that the polluted WDs exhibit a large range in apparent mineralogies that would not be found on Earth. 

\begin{figure}
\vspace{-17mm}
\begin{center}
\centerline{\includegraphics[width=4.7in]{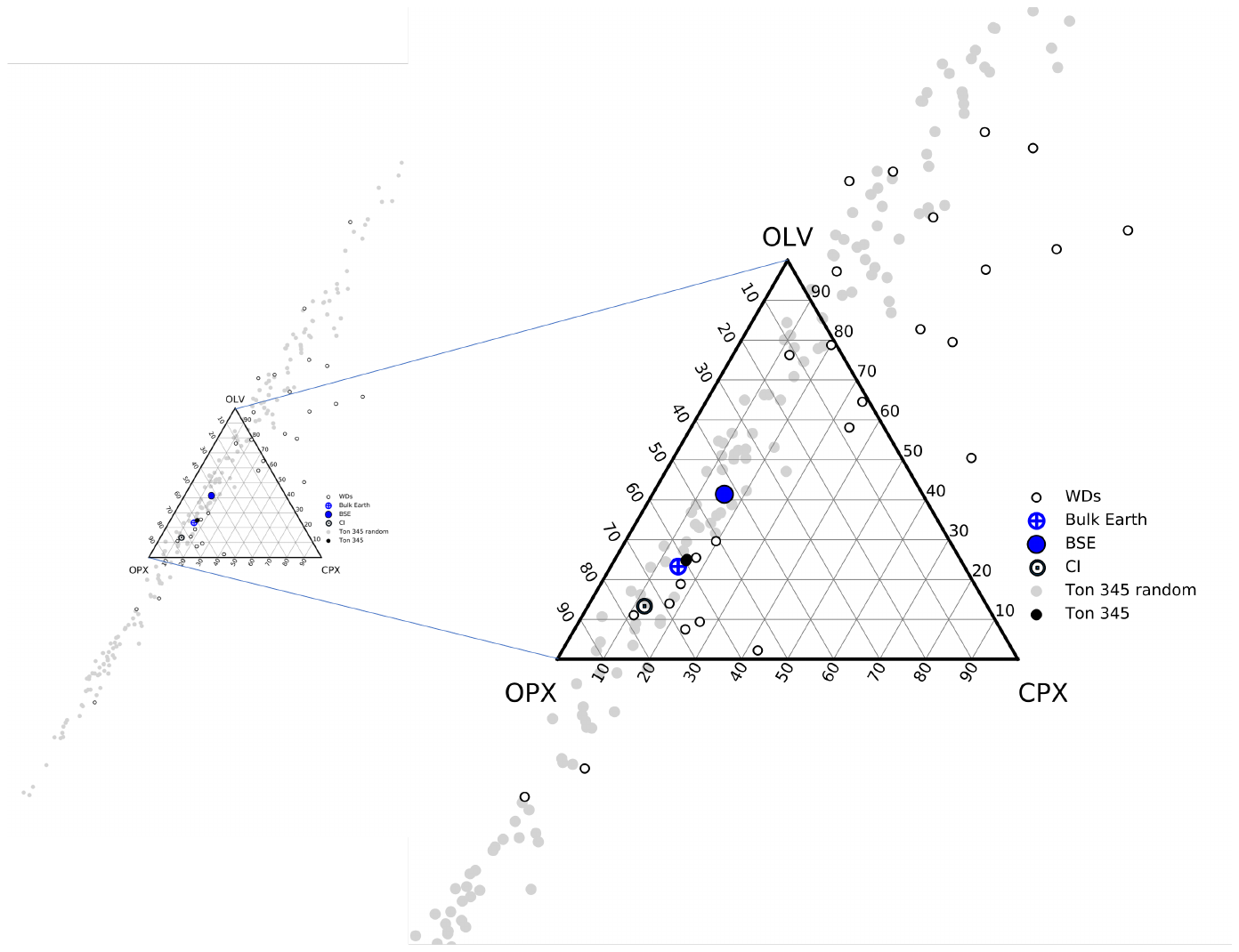}}
\end{center}
\vspace{-14mm}
\caption{Normative mineralogies (zoomed-out to the left, zoomed-in on the right) calculated for various terrestrial reservoir, including bulk silicate Earth (BSE), and CI chondrites, compared with data for the same white dwarfs (WDs) shown in Figures \ref{f.wds1} and \ref{f.wds2}. OLV = Mg$_2$SiO$_4$, OPX = Mg$_2$Si$_2$O$_6$, and CPX = CaMgSi$_2$O$_6$. Also shown are 200 points representing a Monte Carlo simulation of errors associate with the data for WD Ton 345. Note that points can fall outside the ternary plot as a result of the projection algorithm that involves differences in element abundances (e.g., OLV = Mg-Si+Ca).}
\label{f.modes}
\end{figure}

A plot similar to that employed by \citet{Putirka_Xu_2021} for the 31 white dwarfs in Figures \ref{f.wds1} and \ref{f.wds2} is shown in Figure \ref{f.modes}. The ternary plot shows fictive relative volumes of Mg-endmember olivine (OLV, Mg$_2$SiO$_4$), orthopyroxene (OPX, Mg$_2$Si$_2$O$_6$), and clinopyroxene (CPX, CaMgSi$_2$O$_6$), a common plot for terrestrial mantle (ultramafic) rocks. Plotting positions are obtained using a transformation of components such that moles of OLV $\rm = -Si+Mg+Ca$, moles of OPX $\rm = Si-(1/2)Mg-(3/2)Ca$, and moles of CPX $\rm = Ca$. These mole numbers are then converted to relative volumes using molecular weights and molar volumes for the Mg-endmember minerals. The fictive mineral volumes are then normalized to unity for plotting in the ternary space. These plotting positions correspond to the Fe-free system, and represent a section through composition space, as opposed to the alternative of projecting from a single bulk iron composition \citep[e.g.,][]{Putirka_Xu_2021}. This difference leads to changes in plotting positions in the ternary diagram but does not affect the overall spread in data.

The most distinctive feature of the ternary plot in Figure \ref{f.modes} is the large spread in OLV relative to OPX at approximately constant CPX. This spread has been interpreted to mean there are large variations in silicon saturation represented by the rocks consumed by the WDs that are ``exotic" relative to terrestrial rocks \citep{Putirka_Xu_2021}. While some of those ``exotic" rocks varied from terrestrial at the 99\% confidence interval, it is prudent to investigate the application of error to the data. To illustrate this point, a Monte Carlo error analysis for a single WD, Ton 345, composed of 200 random draws from the parent distribution for the reported measurement errors, is also shown in Figure \ref{f.modes}. The spread due to errors associated with the Ton 345 data replicates the spread exhibited by the WD population as a whole, indicating that the large variation in OLV/OPX at nearly fixed CPX may be an artefact of the measurement errors rather than a signature of variable silicon saturation. It is therefore possible that meaningful normative mineralogies cannot yet be calculated from polluted WD data due to significant errors. The small range in normative clinopyroxene, CPX, represents the relative fidelity of the Ca data for which many lines are usually available. 

Based on the ensemble of polluted WDs with sufficient data to characterize the major element composition of the polluting rocks, it appears that rocks around the progenitor A- and F-type stars for these WDs are essentially similar to CI chondrite, or bulk Earth, given the uncertainties. This conclusion is consistent with the studies of chondrite-like oxidation states of rocks polluting WDs \citep{Doyle19} and inferences from polluted WDs that the initial complement of the short-lived radionuclide $^{26}{\rm Al}$ in our Solar System was also not unusual \citep{jura2013al26}. Evidence for rocks that would be considered ``exotic" relative to Solar System rocks is still on-going.

\cite{Trierweiler_2023} applied the same approach outlined above for white dwarfs to the Hypatia Catalog stars for comparison. Their results for the Hypatia Catalog stars are similar to the analysis of white dwarf data in that for the relative abundances of the elements Mg, Si, Ca, Al, Fe, Cr, and Ni, 75\% of the Hypatia Catalog stars are statistically indistinguishable from chondrites.  A truncated sampling of these stars, considering only those that fall within 150 pc of the Solar System, yields a similar result, with 74\% of those stars passing as chondritic in rock-forming element ratios.  For those stars that are not chondritic, the authors found that the siderophile elements are typically lower than the lithophile elements by a factor of about 2.  The explanation lies with the Galactic chemical evolution of our Galaxy.  The ratio of major siderophile to lithophile elements varied with time in the Galaxy as a result of late addition of Fe, Cr, and Ni due to the delayed influence of Type Ia supernovae.  Type Ia supernovae are the result of exchange of mass in a binary system where one star is a white dwarf, requiring stars in the Galaxy to have evolved for at least hundreds of millions of years prior to their first appearance. In contrast, the major lithophile elements Mg, Si, Ca, and Al are produced by Type II core-collapse supernovae. Because the progenitors of Type II core-collapse supernovae are more massive, and thus short lived, the growth of these elements in the Galaxy was more uniform over time. As a result of this Galactic chemical evoluton, older stars will have a relative paucity of siderophile elements.  The implication is that the metal cores of the oldest planets should have comprised less of the mass of their respective planets compared with Earth's core, on average \citep{Trierweiler_2023}. 

\section{Summary}

The detection of small planets allows us a chance to understand the context in which the Earth was formed and the extent to which it, and subsequent life on the surface, is unique. While measuring planetary mass and radius provides a bulk density, there are a variety of more detailed interior structures that can reproduce the planetary density. Stellar elemental abundances, especially for refractory elements such as Mg, Si, and Fe, are therefore a very useful solution to help break the degeneracies, since stars and planets are formed from the raw material within a molecular cloud. However, there are additional physical and chemical processes that may have influenced the formation of a planet, creating a discrepancy between stellar and planetary make-up. In these cases, a planet may appear less dense (like a super-Earth or mini-Neptune) or more dense (similar to a super-Mercury) than predicted from stellar abundances. These distinctions are compounded by the uncertainties in the stellar and planetary properties, the lack of mass measurements for most (small) planets, and the variation in stellar abundances are reported by different techniques. 

Fortunately, the accretion of extrasolar rocks (e.g., asteroids, comets, moons, or planets) onto the surface of WDs provides a unique insight into the composition of small planets not afforded by other means. The ratio of lithophile elements to siderophiles from polluted WDs suggest that extrasolar rocks in the solar neighborhood are similar in overall chemistry to those in the Solar System. However, the fractions of metal cores, rocky mantles, and volatiles for small exoplanets may be quite variable.

\section{Acknowledgements}
NRH acknowledges NASA support from grant \#20-XRP20\_2-0125; she would also like to thank Dr. Amílcar R. Torres-Quijano, CHW3, Tatertot, and Lasagna. The research shown here acknowledges use of the Hypatia Catalog Database, an online compilation of stellar abundance data as described in \citet{Hinkel14} that was supported by NASA's Nexus for Exoplanet System Science (NExSS) research coordination network and the Vanderbilt Initiative in Data-Intensive Astrophysics (VIDA). The results reported herein benefited from collaborations and/or information exchange within NASA's Nexus for Exoplanet System Science (NExSS) research coordination network sponsored by NASA's Science Mission Directorate. This work was supported in part by NASA 2XRP grant No. 80NSSC20K0270 to EDY. EDY acknowledges input from Alexandra Doyle and Isabella Trierweiler (UCLA) in analyzing the white dwarf data discussed herein.



\bibliographystyle{elsarticle-harv}





\end{document}